\documentstyle[prb,aps,psfig]{revtex}
\begin{document}
\title{Spin polarised 
carrier injection into
high $T_c$ superconductors~ :~ A test for the superconductivity mechanism ?}
\author{Satadeep Bhattacharjee and Manas Sardar}
\address{ Materials Science Division, IGCAR, Kalpakkam 603102, India}
%\date{\today}
\maketitle
\begin{abstract}
We point out in this short communique, that be it a s-wave or d-wave
superconductor, in a non equalibrium situation(i.e. in presence of
excess unpolarised/polarised quasi-particles maintained by an
injection current) the superconducting gap suppresion by the presence
of same amount excess quasi-particles, can at best differ by a factor
of 2, for a conventional BCS superconductor.
For the high $T_c$ superconductors on the other hand, there is a
huge difference in gap suppresion
 between upolarised/polarised quasiparticle
injection, as observed in
the experiments\cite{gold}. We argue that this this large anomaly
has a natural explanation
within the {\it Interleyer tunneling mechanism} of Wheatley, Hsu and
Anderson\cite{and}, and is
due to the excess polarised quasiparticles blocking 
the interlayer pair tunneling  process.
We also point out 
that spin polarised quasi-particle injection in a superconductor is a
very easy way to distinguish between a s-wave  or an anisotropic gap
superconductor.

\end{abstract}
\vskip 0.5cm

Spin polarised electron tunneling from
ferromagnetic metals to superconductors
\cite{gold} or strongly correlated metals have
recently generated much interest.
The focus is on trying to understand spin dependent transport
properties of electrons, spin relaxation and possible superconducting
devices.
One is also interested in the
mechanisms of superconductivity suppression(like
reduction in critical currents) due to  tiny
injected tunneling currents(unpolarised/polarised)
across an insulating  junction into the superconductor.

Though original experiments on spin polarised tunneling
in superconductors were carried out in the seventies by
Tedrow and Meservey\cite{ted} and first theoretical attempt was made
by Aronov\cite{aronov}
recently this subject has caught the attention of the physics
community due to discovery of
lanthanam manganite( CMR materials). In these materials the degree of
polarisation of the charge carriers is close to unity(almost perfect
ferromagnetic metal).

Both these experiments\cite{gold} are done in the 3-layer
structure like, HSTC/I/Manganite, where a high $T_c$
superconductor(HSTC) thin film is separated from a ferromagnetic
metallic underlayere(Manganite) by an insulating(I) layer.
The critical current of the HSTC film is found to drop
precipitously as a function of a tiny injection current pushed from
the ferromagnetic metal through the insulating junction into the HSTC
film. When Manganite is replaced by Au(an ordinary paramagnetic metal)
, then it is found that for the
same amount of injected current the drop in the critical current
of the HSTC film is much smaller.
The fractional
change in critical current for polarised and
unpolarised injection currents( over the same
injection current interval)differs by an
order of magnitude.

 The mechanism determining the critical current of a superconducting
film is very complex. However critical current suppresion should be
related to the order parameter suppresion and the puzzle is why 
spin polarised injected carriers are so damaging to superconductivity
in the high $T_c$ materials.
\par
In this paper we try to 
show that the large difference between polarised/unpolarised
injection current is very special only to the high $T_c$
superconductors. We predict that if the high $T_c$ superconductor is
replaced by an ordinary BCS superconductor then the relative
difference in superconducting gap suppresion
in the two cases is not at all that striking(differs by at most a
factor of 2 !!). This is our main result.
In other words spin polarised electron tunneling
into high $T_c$ superconductors can be used to
understand the mechanism of superconductivity in the high $T_c$ materials.
\par
The suppression of superconducting gap due to
unpolarised quasiparticle overpopulation( a
non-equilibrium situation) was first investigated
theoretically  by Owen and Scalapino 
\cite{sca}.
They examined a simple model
of an electron gas containing both Cooper pairs
and excited quasi-particles with the ratio of paired to unpaired
electrons being artificially specified, instead of being uniquely
determined by the temperature, as is usual, in the thermal equilibrium
situation for an isolated superconductor.
 The system is considered as being in thermal equilibrium
although the paired and unpaired electrons are not in chemical
equilibrium. It is assumed that for small amount of excess injected
carriers, the time for pair recombination is much larger that the time
for the injected quasiparticles to thermalize with the lattice. This
is achieved by introducing a chemical potential for the quasiparticles
different from the pair chemical potential.

The free energy of the superconductor $F_s$ is given by,
\begin{equation}
F_s= 2\sum_k \vert \epsilon_k-\mu)\vert (f_k -2f_kv_k^2 +v_k^2)
-\sum_{k,k^{\prime}}V_{k,k^{\prime}}u_kv_ku_{k^{\prime}}v_{k^{\prime}}
(1-f_k)(1-f_{k^{\prime}}) \linebreak[5]
-T\sum_k \lbrack f_k{\rm log}f_k +(1-f_k){\rm log}(1-f_k) \rbrack
\end{equation}
where $u_k~{\rm and}~v_k$ has their usual meaning, and $f_k=\langle
\gamma_{k,\sigma}^{\dagger}\gamma_{k,\sigma} \rangle $ is the
fermi function for the superconducting quasiparticles and $\epsilon_k$
is the dispersion of the electrons in the normal state.
However in addition to the usual constrain equation for the total
number of electrons $\sum_{k,\sigma}\langle c_{k,\sigma}^{\dagger}
c_{k,\sigma}\rangle=N$ which is enforced by  proper choice of the
chemical potential $\mu$, an additional constrain on quasi-particle
excitation number will be imposed, $\sum_{k,\sigma}f_{k,\sigma}=n<<N$.
This can be done by introducing an extra chemical potential\cite{koll} $\mu^*$
so that now, $f_k=[1+e^{\beta(E_k-\mu^*)}]^{-1}$.
The modified BCS gap equation is now,
\begin{equation}
\Delta_k=\sum_{k^{\prime}}^{\prime}{ V_{k,k^{\prime}}\Delta_{k^{\prime}}
\over 2E_{k^{\prime}}}
(1-f_{k^{\prime}\uparrow}-f_{-k^{\prime}\downarrow})
~=~\sum_{k^{\prime}}^{\prime}{ V_{k,k^{\prime}}\Delta_{k^{\prime}}
\over 2E_{k^{\prime}}}
{\rm tanh}\beta/2 (E_{k^{\prime}}-\mu^*)
\end{equation}

Asuming a momentum independent $V_{k,k^{\prime}}$( s-wave
superconductor) and going from
monentum summation to  energy integration with a cut-off $\omega_D$, we get,
$$
{1\over N(E_F)V}=\int_{-\omega_D}^{\omega_D}{d\epsilon \over 2E} {\rm
tanh}{1\over 2}\beta(E-\mu^*)
$$
The number of excess quasi-particles is,
\begin{equation}
n= 2\sum_k[f(E_k-\mu^*)-f(E_k)]=4N(E_F)\int_0^{\infty} {d\epsilon
\over e^{\beta(E-\mu^*)}+1}
\end{equation}
the factor of 2 before the momentum summation is because of
two spin species.

Defining $n_0={n\over 4N(E_F)\Delta_0}$, where $\Delta_0$ 
s the gap at $T=0$ and $\mu^*=0$, we can easily solve for the gap
value at any $T$ and $n$ from the above two equations.
 In the limit of zero temperature 
the gap value can be determined from the algebraic equation,
\begin{equation}
{\Delta_0 \over \Delta}= \lbrack n_0 {\Delta_0 \over \Delta} ~+~\sqrt{1+
n_0^2 {\Delta_0^2 \over \Delta^2}} \rbrack^2
\end{equation}
This is the result of Owen and Scalapino, for excess unpolarised 
 quasiparticles.

For spin polarised quasiparticle injection, we 
introduce a different chemical potential, for
{\it only the up spin quasiparticles}, i.e. assuming complete spin
polarisation of electrons in the ferromagnet. The gap equation (2)
will be modified to,
\begin{equation}
\Delta_k=\sum_{k^{\prime}}^{\prime} {V_{k,k^{\prime}}\Delta_{k^{\prime}}
\over 2E_{k^{\prime}}}{1\over 2}\lbrack 
{\rm tanh}\beta/2 (E_{k^{\prime}}-\mu^*)+{\rm tanh}\beta/2
E_{k^{\prime}}\rbrack
\end{equation}
Corresponding equation for the number of excess quasiparticles will
be, same as equation (3) with the factor of 2 missing before the
momentum summation, because excess quasiparticles are of only one
spin species. In the limit of zero temperature we get the folowing
algebraic equation for the normalise gap value,
 
\begin{equation}
{\Delta_0 \over \Delta}= \lbrack 2n_0 {\Delta_0 \over \Delta}
~+~\sqrt{1+
4n_0^2 {\Delta_0^2 \over \Delta^2}} \rbrack^2
\end{equation}

In figure.1 we show the normalized
gap versus extra quasiparticles for  a s-wave
superconductors(solutions of equations 3 \& 4). We find that for, s-wave
superconductors, a first order phase transition to normal metal
occurs, for $n=0.15;~ \Delta= 0.63\Delta_0$, and 
$n=0.08;~ \Delta= 0.58\Delta_0$ for unpolarised
and polarised quasiparticle injections. Beyond this critical
concentration of injected carriers, the free energy of the perturbed
superconductor becomes larger than normal state free energy. 
Notice though that, for the same amount of injected carriers, the
relative gap suppresion in unpolarised/polarised carriers differs by
at most by a factor of 2.

Just as at finite temperatures quasi-particle excitations interfere
with the pairing interaction and eventually destroy superconductivity
at $T_c$, when an excess number of quasi-particles are injected into
the superconductor, it basically reduces the phase space for BCS pair
scattering process and reduce the gap value.
BCS interaction scatters pairs
$(k\uparrow,-k\downarrow)->(k^{\prime}\uparrow,-k^{\prime}\downarrow)$
accross the fermi surface.
So any excess quasiparticle occupying these states limits the phase
space for the BCS interaction. It is obvious that when the injected
quasi-particles are polarised(all of one spin) they interfere with BCS
interaction more severly 
and hence the gap value falls faster with quasi-particle over
-population, compared to the unpolarised injection current.
\par
We next investigate the effect of quasi-particle injection  on
superconductors with anisotropic gap, specifically gaps of d-wave
symmetry.
The dispersion of electrons is chosen
to be of the form
\begin{equation}
\epsilon (k) ~~ = ~~ -2t~(~cos k_x~+cos k_y~)~~+~~
4t^{\prime}~cosk _x~cos k_y~~-2t^{\prime \prime}(\cos 2k_x+\cos 2k_y),
\end{equation}
with $t$ = 0.25 eV ,${t^{\prime} \over t}$ = 0.45 $, {t^{\prime
\prime}\over
t}=0.2$. We also choose
$\epsilon_F$ = -0.45 eV corresponding to a Fermi surface which is
closed
around the $\Gamma$ point. 
These choices
are inspired by band structure calculations \cite{band} for
the YBCO compound at optimal doping concentration. 
The cut off the pairing interaction is chosen to be $\omega_D=30 ~meV$,
and  the pairing interaction strength
$V_{k,k^{\prime}}=V_0f_kf_{k^{\prime}}$ with $f_k=\cos(k_x)-\cos(k_y)$
and $V_0=2.8 ~meV$. With this choice of parameters the $T_c$ comes out
to be $30^0K$.

In figure.2 we plot the normalised momentum averaged
 value of the superconducting
gap magnitude, $\Delta(n,T)/\Delta(0,T)$ versus $n$(quasi-particle over
population) for different temperatures.
There are some interesting features to be noticed here.

(1) Notice the crossing of curves, for $T=5~{\rm and}~25^0K$.
The origin of this can be seen by looking at the gap equation(2).
States of energy $E_k$ less that $\mu^*$ interferes with the pairing
process by giving a negative contribution to the binding.
At low temperatures and small injection currents, the extra carriers
occupy the states near the deep gap nodes, where the denominator $E_k$
is also small giving rise to large reduction in binding energy.
At larger temperatures(for same amout of injected carriers)
 on the other hand the quasi-particles are distributed over a larger
range of energies(larger denominator) leading to smaller suppresion
of binding(and hence larger gap values). Larger concentration of 
injected carriers, of course will ultimately be more damaging to
superconductivity, because of larger number of thermally excited
quasi-particles that are already present. This leads to the crossing
of curves as seen in  fig 2. This will be a generic feature for any
superconductor with deep gap nodes, and should be easily seen in spin
injection experiments.

(2) Notice also  that the critical concentration of excess
quasi-particles that destroy superconductivity is not a monotonic
function of temperature and goes through a maximum around $T_c/2$.

(3) Though it is not shown in the plot, injected current suppreses
superconductivity more in the d-wave superconductors, than for a
s-wave
superconductor having same critical temperature,
as one would expect for gap
functions with deep nodes. 

In figure.3, we have plotted the normalised momentum averaged gap
magnitude of a d-wave superconductor, with a $T_c$ of $90^0K$ versus
for both unpolarised/polarised
quasi-particle over population( we go upto $n=0.15$ only).
The difference in the reduction of gap
values in the unpolarised/polarised tunneling is
the range of $2.5-5.0\%$(i.e a factor of 2 only). 
Clearly the large anomaly
observed  while tunneling into the high $T_c$
superconductor, as seen in the experiments\cite{gold}
 has its origin in  the superconducting
mechanism in the high $T_c$ materials itself.

We shall explore here the
{\it interlayer tunneling model} of
superconductivity\cite{and,chak}.
\par
We begin by writing the gap equation for interlayer tunneling model
 with unpolarised injected carriers as,
\cite{chak},
\begin{equation}
\Delta_k  =  T_J(k) {\Delta_k \over 2E_k}
{\rm tanh} {\beta (E_k-\mu^*) \over 2} + ~ \sum_{kk^{\prime}}^{\prime}
V_{kk^{\prime}}~
{\Delta_k^{\prime} \over 2E_{k^{\prime}}} {\rm tanh}
{\beta (E_{k^{\prime}}-\mu^*) \over 2},
\end{equation}
For polarised carrier injection, this equation has to corrected as
shown in equation (5).

This gap equation can be obtained by considering two close Cu-O layers as
in
YBCO  coupled by a Josephson tunneling term of the form
$$
H_J~~ = ~~ -{1 \over t}
\sum_{k}~~t^{2}_{\perp}(k)~~(~c^{\dag}_{k\uparrow}
c^{\dag}_{-k\downarrow} d_{-k\downarrow} d_{k\uparrow} + {\rm
h.c.}~)~~,
$$
where $t_{\perp}(k)$ is the bare single electron hopping term between
the two coupled layers $c$ and $d$ and $t$ is a band structure
parameter
in the dispersion of electrons along the Cu-O plane.  Finally,
$T_J(k)$
in the right hand side of equation (1) is given by $T_J(k) =
{t^2_{\perp}(k)
\over t}$.
where $t_{\perp}(k)={t_{\perp}\over 4}({\rm cos}(kx)-{\rm
cos}(ky))^2$.
The dispersion of electrons along the Cu-O plane is is given by
equation 5.
Note that
the
Josephson coupling term in $H_J$ conserves the individual momenta of
the
electrons that get paired by hopping across the coupled layers.  This
is
as opposed to a BCS scattering term  which would only conserve the
center of mass momenta of the pairs.  This is the origin of all
features
that are unique to the interlayer tunneling mechanism.
This term has a local $U(1)$ invariance in $k$-space and cannot by
itself give a finite $T_c$. It needs an additional BCS type non local
interaction in the
planes which could be induced by phonons or residual correlations.
Here we
assume the inplane pairing interaction to be d-wave kind i.e,
$V_{kk^{\prime}}=V_0~f_kf_{k^{\prime}}$
with $f_k=\cos k_x -\cos k_y$. $T_J(k)$ can be infered from electronic
structurecalculations. As shown in reference\cite{band}, it is
adequate to choose $t_{\perp}(k)=
{t_{\perp}^2\over 4}(\cos k_x -\cos k_y)^2$, with $t_{\perp}/t\equiv
1/3~{\rm to}~1/5$.
According to Anderson, it is the $k$-space locality that leads to a
scale of
$T_c$ that is linear in interlayer  pair tunneling
matrix element. He finds that in the limit $T_J > V_{kk^{\prime}}$,
$k_BT_c
\approx {T_J\over 4}$ and in the other limit, $k_B T_c\approx \hbar
\omega_D
e^{-{1\over \rho_0 V_0}}$, where $\omega_D$ $\rho_0$ and $V_0$ are
Debye
frequency, density of states at the fermi energy, and fermi surface
average matrix
element $V_{kk^{\prime}}$.
The important point being that even with a little help from the in
plane
pairing interaction the interlayer tunneling term can provide a large
scale
of $T_c$.
In this model the gap value is mainly controlled by the interlayer
tunneling amplitude, rather than the inplane
pairing interaction,
 though the symmerty of the gap function is
determined overwhelmingly by the inplane BCS kind of interaction
strength $V_{k,k^{\prime}}$.
 and the gap value is very
sensitive to the interlayer pair tunneling
amplitude\cite{manas}.

In figure.4, we plot the normalised momentum averaged gap
values versus both unpolarised/polarised 
quasi-particle over population upto $n=0.085$.
We find that for $n=0.085$, the reduction in gap value
for unpolarised injection current is about 5\%, whereas for the
polarised carrier injection is about 35\%( that is a factor of 7 !).
\par
The large difference between the two situations(as observed in the
experiments\cite{gold} also) now shows up.
 We argue that when polarised carriers are injected,
then, over and above the usual
dynamical pair breaking(due phse space blocking
of BCS interaction)in the planes like in
usual BCS superconductors, there is an added
inhibition of  interlayer pair tunneling.
This is
so, because there are much less number of
 extra singlets near the fermi surface, to tunnel from plane to plane.
This  effect is absent when unpolarised
quasiparticles enter the planes.
In this mechanism ,{\it  extra quasiparticles directly
affect the interlayer pairing tunneling process, which
is the main source of superconducting condensation energy gain}.
This is over and above the interference in the binding process
in the  individual layers that  we have discussed earlier
in usual BCS superconductors.
\par
The introduction of an extra chemical potential $\mu^{\star}$
to tackle the nonequilibrium superconductivity, in presence of
artificially maintained quasiparticles, is a reasonable starting
point,
when the excess quasiparticles thermalize with low energy phonons more
often than they recombine into pairs. In these limit the
quasiparticles are in steady state at $T$ but have an excess number
denoted by the increased chemical potential $\mu^{\star}$. 
One important ingradient of Anderson et al's\cite{and} mechanism
is that, in the normal state the electrons are spin-charge separated
and no quasiparticles in the fermi liquid sense exists.
We have assumed here that,
below the superconducting
Tc somehow spin-charge separation is absent. This seems to be the
case from photoemission experiments which show clear signal
of well defined quasiparticles on the fermi surface.In the superconducting
state how exactly it happens, is still not clear. 

  At the present moment all that we can say is that,
because of spin-charge separation in the normal state, the
single particles(which are not even well defiend properly) cannot
tunnel from plane to plane in a coherent fashion.
On the other hand there are indications from some
earlier theoretical attempts\cite{muthu}
that electrons can tunnel in pairs.
In other words spin-charge separation itself is responsible for
pair tunneling.
We believe that when some BCS type of
pairing interaction is introduced, this anomolous self energy
actually leads to weaking of spin charge separation, and
electrons recover fermi liquid type quasi-particles below Tc,
as it is observed in the photoemission experiments.
We have implicitly assumed this while working with
the interleyer tunneling mechanism. Clearly a more carefull analysis
of this problem is needed within this mechanism.

\par
Throughout this analysis, we have assumed that,
(1) the London penetration depth is lesser that
the superconducting film thickness, so that direct
magnetic field of the ferromagnetic metal does not
penetrate the superconductor much.
(2) The spin diffusion length is much larger that
the SC film thickness, so that no spatial
variation of the gap has to be taken into account.
This is certainly true when there are no magnetic
impurity or for small spin orbit interaction, for
the then extra spin density  relaxes slowly.

We have also not taken into account the case of
finite recombination time for the excess
quasiparticles.

It would be interesting to see if, the presence of extra spin polarised
quasiparticles will give rise to polarisation
field, that will be felt by the nuclear magnetic
moment through the contact hyperfine interaction.
This will significantly affect the  NMR relaxation
rate and ESR linewidths.
 
\par
Recently  N. C. Yeh et al\cite{yeh},have observed
an initial, actual increase in critical
current for low enough injection currents, when the insulating barrier
thickness is small.
They argue that, some up spin quasiparticles in the superconductors
can diffuse into the magnetic materials, creating spin imbalance
in the superconductor(more of down spin quasiparticles). On the other
hand when the injection current is switched on, then up spin
electrons starts coming into the superconductor, nullifying the spin
imbalance in the superconductor. That is why Tc increases for small
injection currents. For larger injection currents of course Tc falls
drastically as  is seen in experiments, and as we have argued to be
natural within the Interlayer tunneling mechanism of Anderson et al.

While this maybe the case, we would also like to
point out, that this initial increase
can also happen, when there are some (1) magnetic impurity
inside the superconductor or (2) spin-orbit interaction is important.
Because the injected polarised carriers will give an internal
magnetic field which will polarise the magnetic impurities,
so that, they become less effective in breaking pairs.
Spin orbit interaction can similarly be counteracted by the internal
magnetic field due to spin imbalance in the material, though the spin
orbit interaction might be absent in the cuprate superconductors.

\vskip 2.5cm

\begin{figure}
\caption{ $\Delta (T=0,n_0)/ \Delta(T=0,n_0=0)$ 
versus $n_0$( the quasi-particle over population, as defined in the
text) for s-wave superconductor(the  solutions of equations 3 \& 4)}
\label{Fig.1}
\end{figure}

\begin{figure}
\caption{The normalised momentum averaged gap
$\Delta(T,n)/ \Delta(T,n=0)$ of a d-wave superconductor, with
$T_c=30^0{\rm K}$ versus injected carrier concentration. 
Solid line is for $T=5^0{\rm K}$, long dashes $T=10^0{\rm K}$, short dashes
$T=15^0{\rm K}$, dotted line $T=20^0{\rm K}$ and the dash-dotted line os for
$T=25^0{\rm K}$.}
\label{Fig.2}
\end{figure}

\begin{figure}
\caption{The normalised momentum averaged gap
$\Delta(T,n)/ \Delta(T,n=0)$ of a d-wave superconductor, with
$T_c=90^0{\rm K}$ versus injected carrier concentration at $T=10^0{\rm K}$.
The solid and dashed lines are for unpolarised and polarised
injected carriers.}
\label{Fig.3}
\end{figure}

\begin{figure}
\caption{The normalised momentum averaged gap
$\Delta(T,n)/ \Delta(T,n=0)$ of a (d-wave superconductor~+~
interlayer tunneling), with
$T_c=90^0{\rm K}$ versus injected carrier concentration at $T=10^0{\rm
K}$.
The solid and dashed lines are for unpolarised and polarised
injected carriers.}
\label{Fig.4}
\end{figure}
\newpage

%-------------------------------------------------------------
%\begin{figure}
%\centerline{\psfig{figure=fig1.ps,height=16cm,width=16cm}}
%\end{figure} 
%\newpage
%\begin{figure}
%\centerline{\psfig{figure=fig2-u.ps,height=16cm,width=16cm}}
%\end{figure}
%\newpage
%\begin{figure}
%\centerline{\psfig{figure=fig3.ps,height=16cm,width=16cm}}
%\end{figure}
%\newpage
%\begin{figure}
%\centerline{\psfig{figure=fig4.ps,height=16cm,width=16cm}}
%\end{figure}
%-------------------------------------------------------------

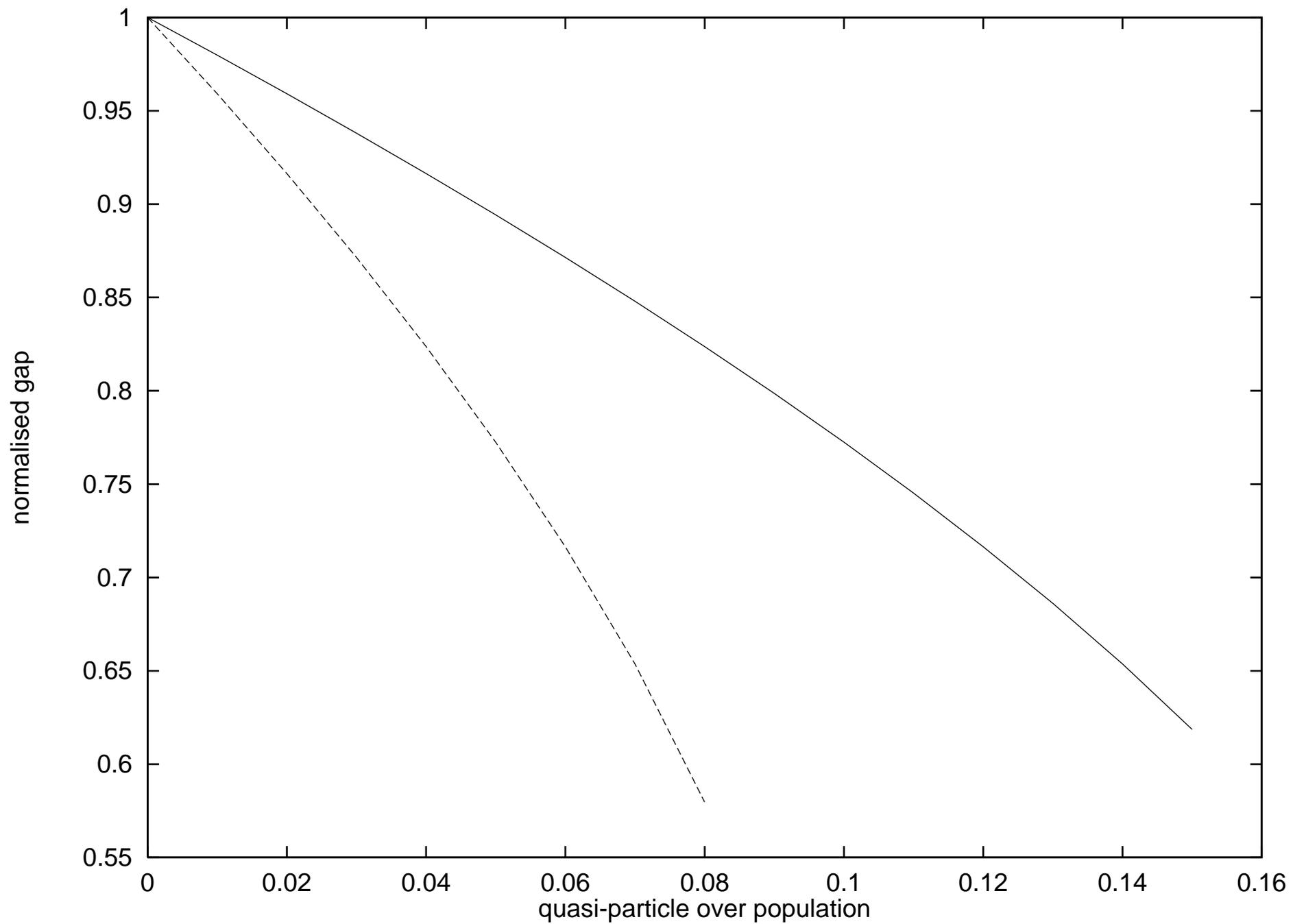
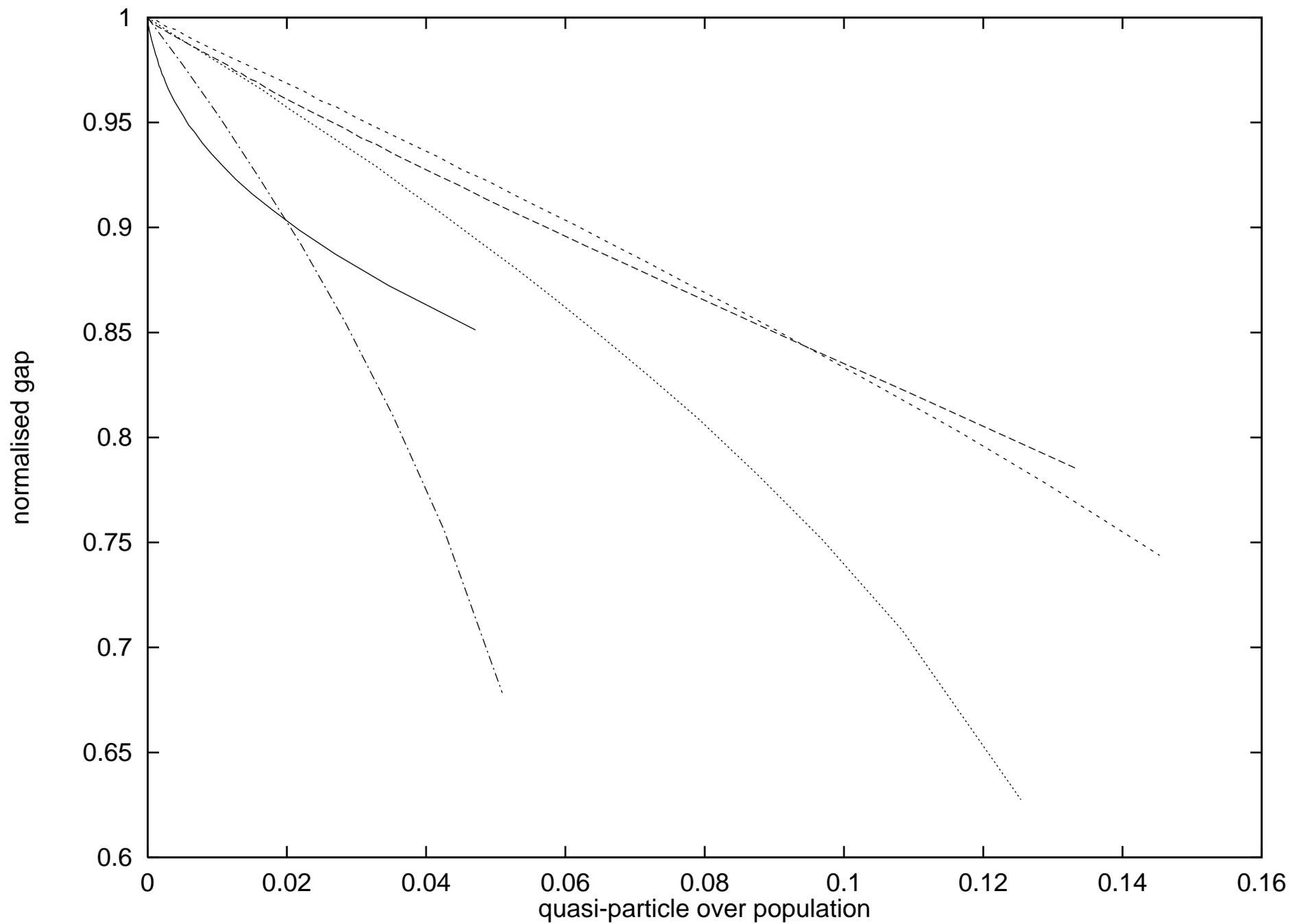
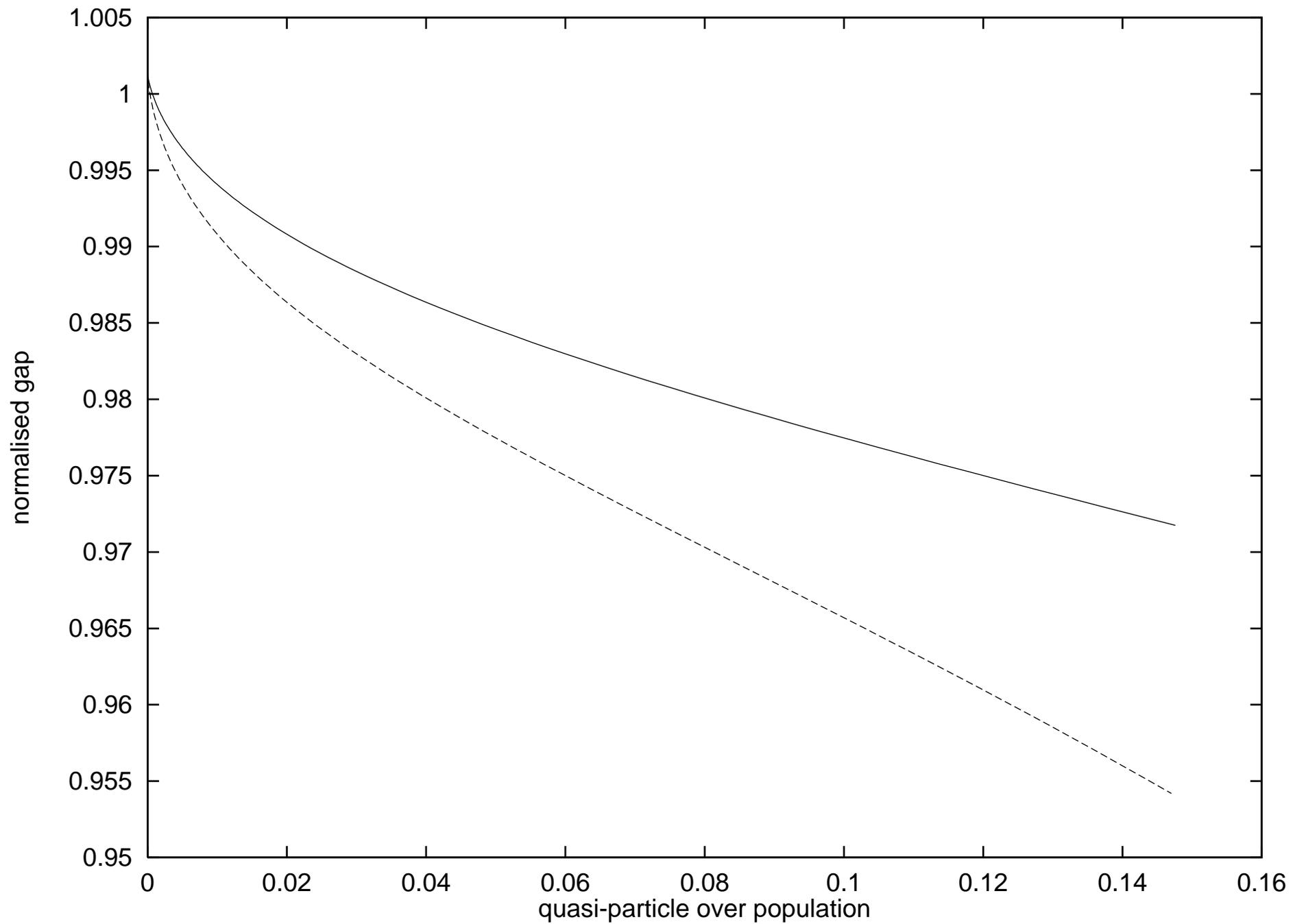
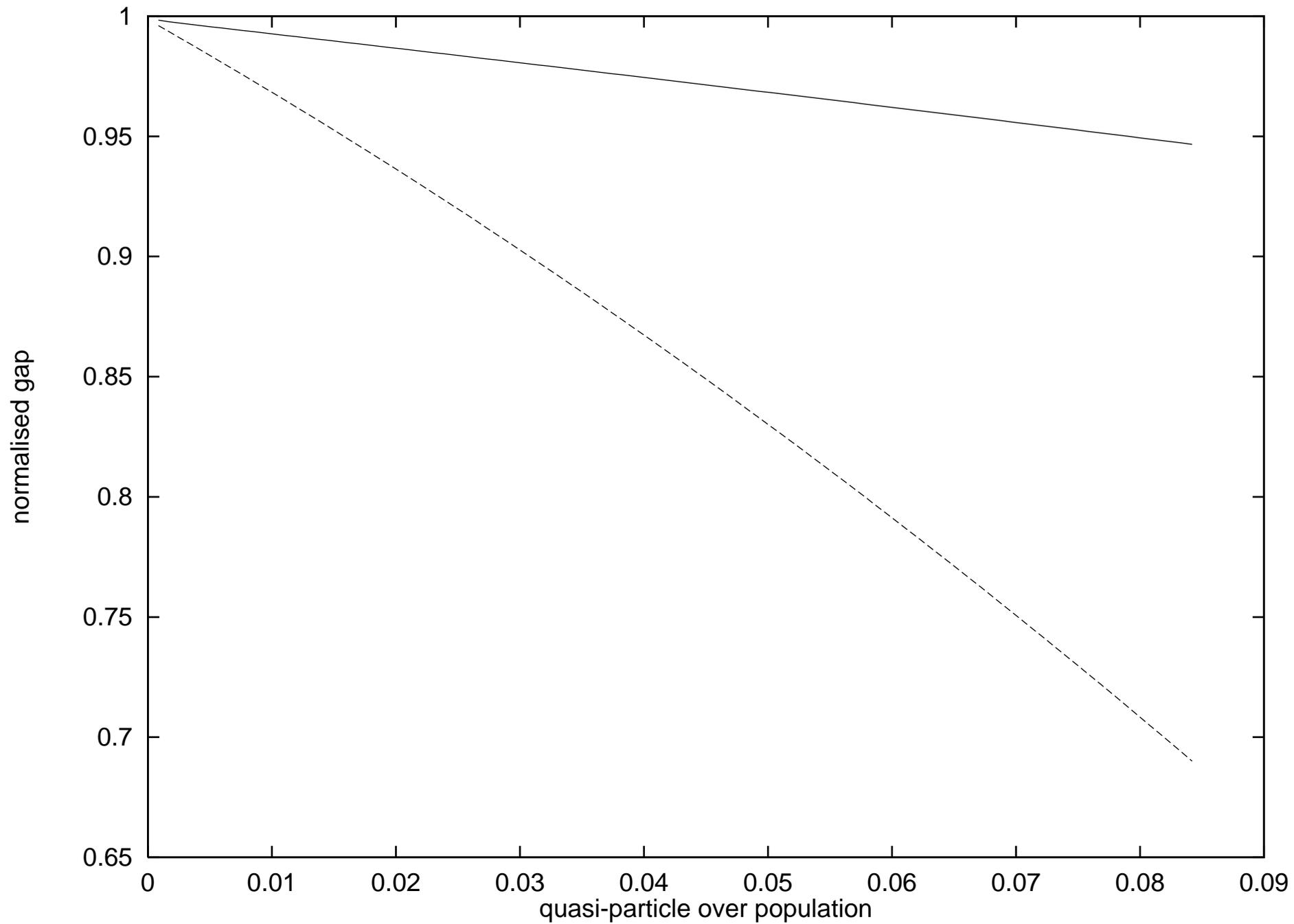
\end{document}